\documentclass[runningheads]{llncs}
\usepackage[T1]{fontenc}
\usepackage{pdfpages,multirow,ragged2e} %
\usepackage{graphicx}
\usepackage{circuitikz}
\usepackage{multirow}

\usepackage{tikz}
\usetikzlibrary{calc}
\usetikzlibrary{automata, arrows,positioning,datavisualization,shapes.geometric}
\definecolor{verylightgray}{rgb}{0.92,0.92,0.92}
\usepackage{hyperref}
\usepackage{amsmath}
\usepackage{orcidlink}
\usepackage{enumitem}
\newcommand{\subscript}[2]{$#1 _ #2$}
\usepackage{float}
\usepackage{wrapfig}
\usepackage{listings}
\usepackage{multicol}
\usepackage{cleveref}

\usepackage{tikz-timing}[2014/10/29]
\usetikztiminglibrary[rising arrows]{clockarrows}
\usepackage{xparse} 

\newif\ifextended
\extendedtrue 

\lstdefinelanguage[Siemens]{ST}{
		morekeywords=[0]{
			ORGANIZATION\_BLOCK, END\_ORGANIZATION\_BLOCK, FUNCTION\_BLOCK, END\_FUNCTION\_BLOCK, FUNCTION, END\_FUNCTION, DATA\_BLOCK, END\_DATA\_BLOCK,
			VAR, VAR\_TEMP, VAR\_INPUT, VAR\_IN\_OUT, VAR\_OUTPUT, END\_VAR,
			STRUCT, END\_STRUCT, LABEL, END\_LABEL, CONST, END\_CONST, TYPE, END\_TYPE,
			BLOCK\_DB, BLOCK\_FB, BLOCK\_FC, BLOCK\_SDB, BLOCK\_SFB, BLOCK\_SFC,   
			BEGIN, EXIT, GOTO, CONTINUE, RETURN, 
			ARRAY, OF, 
			AND, OR, XOR, MOD, DIV, NOT,
			TRUE, FALSE, OK, EN, ENO, RET\_VAL,
			FOR, TO, BY, END\_FOR,
			WHILE, DO, END\_WHILE,
			REPEAT, UNTIL, END\_REPEAT,
			CASE, END\_CASE,
			IF, THEN, ELSE, ELSIF, END\_IF,
            ASSERT
			},
		morekeywords=[1]{ 
			BOOL, INT, UINT, SINT, USINT, DINT, CHAR, BYTE, WORD, DWORD, REAL, TIME, DATE, TIME\_OF\_DAY, DATE\_AND\_TIME, ANY, VOID, POINTER, S5TIME, COUNTER, TOD, DT, STRING, NIL, TIMER},
		sensitive=false,
		morecomment=[s]{(*}{*)},%
		morecomment=[l]//\  ,%
		morestring=[d]'%
	}
\lstalias[]{SCL}[Siemens]{ST}
\lstdefinestyle{assertion}{
  language=SCL,
  breaklines=true,
	captionpos=b,
  showstringspaces=false,
  basicstyle=\footnotesize\ttfamily,
  identifierstyle=\color{black},
  stringstyle=\color{orange},
  columns=flexible,
  keywordstyle=[0]*\bfseries\color{blue!40!black},
  keywordstyle=[1]\mdseries\color{blue!40!black}, 
  commentstyle=\itshape\rmfamily\color{green!40!black},
  frame=tb, 
  numbers=left,
  numbersep=5pt, 
  numberstyle=\tiny\color{gray},
}
\newcommand\blfootnote[1]{%
  \begin{NoHyper}%
  \renewcommand\thefootnote{}\footnote{#1}%
  \addtocounter{footnote}{-1}%
  \end{NoHyper}%
}

\newlist{inlinelist}{enumerate*}{1}
\setlist*[inlinelist,1]{%
	label=\textit{(\roman*)},
}

\NewDocumentCommand{\busref}{som}{\texttt{%
#3%
\IfValueTF{#2}{[#2]}{}%
\IfBooleanTF{#1}{\#}{}%
}}

\usepackage{color}

\begin{document}
\ifextended
\title{Formal Verification of PLCs as a Service: A CERN-GSI Safety-Critical Case Study \ \ \ \  (extended version)}
\else
\title{Formal Verification of PLCs as a Service: A CERN-GSI Safety-Critical Case Study}
\fi

\titlerunning{Formal Verification as a Service: A CERN-GSI Case Study}

\author{Ignacio D. Lopez-Miguel\inst{1}\orcidlink{0000-0002-8044-0385} \and
Borja Fernández Adiego\inst{2}\and
Matias Salinas\inst{3}\and Christine Betz\inst{3}}
\authorrunning{Ignacio D. Lopez-Miguel et al.}

\institute{TU Wien, Vienna, Austria \email{ignacio.lopez@tuwien.ac.at} \and
CERN, Beams Department, Geneva, Switzerland \email{borja.fernandez.adiego@cern.ch}\and
GSI, Darmstadt, Germany
\email{\{m.salinas,c.betz\}@gsi.de}}
\maketitle              
\begin{abstract}

The increased technological complexity and demand for software reliability require organizations to formally design and verify their safety-critical programs to minimize systematic failures.
Formal methods are recommended by functional safety standards (e.g., by IEC 61511 for the process industry and by the generic IEC 61508) and play a crucial role.
Their structured approach reduces ambiguity in system requirements, facilitating early error detection.
This paper introduces a formal verification service for PLC (programmable logic controller) programs compliant with functional safety standards, providing external expertise to organizations while eliminating the need for extensive internal training. It offers a cost-effective solution to meet the rising demands for formal verification processes. 
The approach is extended to include modeling time-dependent, know-how-protected components, enabling formal verification of real safety-critical applications.
A case study shows the application of PLC formal verification as a service provided by CERN in a safety-critical installation at the GSI particle accelerator facility. 

\end{abstract}
\section{Introduction}
\label{sec:introduction}
\ifextended
\blfootnote{This paper is an extended version of our NFM 2025 paper ``Formal Verification of PLCs as a Service: A CERN-GSI Safety-Critical Case Study''. It adds an appendix
with the complete modeling of a know-how-protected function and with examples of found discrepancies during verification.} 
\fi

Formal methods play an essential role in ensuring the reliability, quality, and safety of software systems. They are recommended by industry standards and provide a mathematical approach to software development.
One of these standards is DO-178C~\cite{do178c} in the aviation domain, which is accompanied by a guideline on formal methods (DO-333~\cite{do333}). The latter enhances the former by explaining how to use formal methods in every stage of the software lifecycle. 

Large scientific installations, like particle accelerators, do not have specific standards. However, they tend to apply the generic IEC 61508~\cite{iec61508} and IEC 61511~\cite{iec61511} functional safety standards to design, develop, and validate their safety-critical software. IEC 61508 recommends using formal approaches in different parts of the software lifecycle according to the criticality of the component. IEC 61511 recommends the usage of formal methods to specify requirements.
ISO 26262~\cite{iso26262} for road vehicles also recommends the use of formal methods for critical components, and IEC 61513~\cite{iec61513} for nuclear power plants emphasize the importance of rigorous development and verification processes to ensure the safety and reliability of safety-critical systems.

All these standards agree that, although formal methods can be expensive, identifying discrepancies between the code and the requirements in the early development stages results in substantial cost reduction in later phases. 

However, some organizations might lack the resources to introduce formal methods in their software development process. \textit{Formal verification as a service} addresses this need, offering formal methods expertise. It establishes a win-win situation where organizations benefit from the skills of experts, and service providers improve their tools based on the different case studies. It contributes to quality assurance, enabling organizations to demonstrate to regulatory authorities that exhaustive measures have been taken to ensure safety.

In this paper, we focus on the formal verification of PLC (programmable logic controller) programs as a service. Our contributions are summarized below:
\begin{enumerate}
    \item We present a collaboration model between the different stakeholders of a PLC project development and the formal verification service providers. It complies with the functional safety standards by ensuring independence and by using formal verification at the early stages of the PLC program lifecycle.
    \item We introduce a methodology based on simulation and formal verification to model \textit{know-how-protected functions}, which are proprietary functions whose precise behavior is hidden by the manufacturer (black boxes). They are commonly used in PLC programming, and some include time-dependent components, complicating their modeling. Their exact behavior must be known to formally verify a complete PLC program containing these functions.
    \item We show a real case study in which PLCverif~\cite{plcverifweb} was used to verify a safety-critical system containing know-how-protected functions at the particle accelerator installation at GSI Helmholtz Centre for Heavy Ion Research~\cite{gsi}.
\end{enumerate}

\section{Service approach}
\label{sec:approach}

\subsection{Collaboration model}
\label{subsec:collaboration}

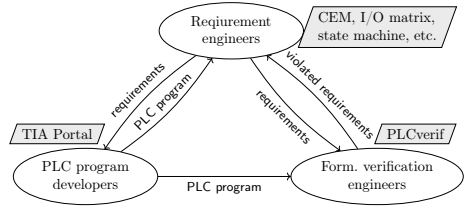
\begin{wrapfigure}{r}{0.55\linewidth}
    
    \resizebox{.5\textwidth}{!}{\begin{tikzpicture}[->, node distance=5cm]

  \node [draw, shape=ellipse, align=center, minimum width=3.5cm] (specification) {Reqiurement\\engineers};  
  \node [draw, shape=ellipse, align=center, minimum width=3.5cm] (plc) [below left of=specification, align=center] {PLC program\\developers};
  \node [draw, shape=ellipse, align=center] (formalVerification) [below right of=specification] {Form. verification\\engineers};
  \node [draw, shape=trapezium, trapezium left angle=70,trapezium right angle=-70, align=center,fill=verylightgray] (formalisms) [above right =-.9cm and 1.8cm of specification, align=left] {CEM, I/O matrix,\\state machine, etc.};
  \node [draw, shape=trapezium, trapezium left angle=70,trapezium right angle=-70, align=center,fill=verylightgray] (tiaportal) [above left =.3cm and -.8cm of plc] {TIA Portal};
  \node [draw, shape=trapezium, trapezium left angle=70,trapezium right angle=-70, align=center,fill=verylightgray] (plcverif) [above right =.3cm and -.8cm of formalVerification] {PLCverif};
    
  \path[every node/.style={font=\sffamily\small}]
    (specification) edge [bend right=10] node [above,rotate=45] {requirements} (plc)
    (plc) edge [bend right=10] node [above,rotate=45] {PLC program} (specification)
    (plc) edge node [below] {PLC program} (formalVerification)
    (specification) edge [bend right=10] node [below,rotate=-45] {requirements} (formalVerification)
    (formalVerification) edge [bend right=10] node [above,rotate=-45, pos=0.44] {violated requirements} (specification)
    	    ;
\end{tikzpicture}}
    \caption{Diagram with the roles of the collaboration, shared information and used tools.}
    
    \label{fig:collaboration}
\end{wrapfigure}

Figure~\ref{fig:collaboration} depicts the proposed diagram to offer formal verification of PLC programs as a service~\cite{lopez:ICALEPCS23}. It is composed of the following independent teams as recommended by IEC 61511-1 clause 12~\cite{iec61511-1} of having an independent body in charge of validating the critical software: 
\begin{itemize}
    \item \textit{Requirement engineers}. They are responsible for analyzing the systems and writing their technical requirements using different formalisms. They have the best knowledge of the actual physical system for which the PLC program is developed, and they know how the system should behave. That is why they write and distribute the requirements to the other teams.
    \item \textit{PLC program developers}. They follow the requirements handed out by the requirement engineers to implement the PLC program. If the requirements are clear, the interaction with the requirement engineers can be minimal. Their PLC program is then shared with the other two teams.
    \item \textit{Formal verification engineers}. They ensure that the PLC program behaves exactly as written in the requirements using formal verification. The requirements engineers are informed when a discrepancy between the PLC code and the requirements is found. Then, they work with the developers to solve it.
\end{itemize}

It is important to highlight the iterative nature of this process. Especially when a discrepancy is found, formal verification engineers need to inform requirement engineers, who will work with the developers to find the root cause of the error. This will lead to updated requirements and/or PLC programs, which are then given again to the formal verification engineers so they can continue their work. This process is repeated until no more discrepancies are found. 

Although this process does not entirely ensure the lack of errors in the code or in the requirements due to the possible bugs in the program verifier, it drastically increases the confidence of the requirement engineers with the PLC program. It also helps them show authorities they made considerable efforts to guarantee the safety of the installation. In fact, formal verification, compared to other methods like testing, can identify more hidden bugs (corner cases).

Another important formal method to mention at this point is the synthesis of PLC programs~\cite{Yoo:plcSynthesis}, which would make formal verification redundant since the code would be correct by construction. However, synthesis tools for PLC programs are not widespread in the industry.

In the next two subsections, from Figure~\ref{fig:collaboration}, we will further explain how the requirement engineers can formalize requirements and how the formal verification engineers verify the given PLC code according to the formalized requirements.

\subsection{Formal specification}
\label{subsec:specification}

Requirements can be represented using diverse formalisms, which should be simple, clear, and concise for use across the software development lifecycle. The examples in this section meet these criteria, were applied in the case study (Section~\ref{sec:caseStudy}), and align with functional safety standards. IEC 61511-2~\cite{iec61511-2} recommends methods like cause-and-effect matrix, state machines, and logic diagrams; IEC 61511-1~\cite{iec61511-1} provides examples of state machines and logic diagrams.

\begin{itemize}
    \item A~\textit{Cause-and-effect-matrix (CEM)}~\cite{adiego:ICALEPCS19} is a tabular representation of Boolean expressions. It is particularly suitable for stateless logic like interlock logic. The example from Table~\ref{tab:CEM} assigns values to the outputs according to $\mathit{Out\_1}=(\mathit{In\_1} \land \mathit{In\_2}) \vee (\neg \mathit{In\_3} \land \neg \mathit{In\_4})$, and $\mathit{Out\_2}=\bigwedge_{i=1}^{4}\mathit{In\_}i$.
    \item An~\textit{input-output matrix}~(I/O matrix) gives the conditions to set or reset output variables. 
    One needs to ensure that the inputs are mutually exclusive or to specify output priorities. The I/O matrix from Table~\ref{tab:IOmatrix} shows an example. 
    \item A~\textit{state machine} is a graphical representation used to depict the behavior of a system consisting of different states and transitions between them.
    Figure~\ref{fig:stateMachine} shows a simple state machine that changes from two modes depending on the requests. For a real example, one can refer to~\cite[Figure~4.4]{nfm2023technicalReport}.
    
\end{itemize}

\vspace{-12pt}
\begin{table}[H]
\centering
\def\arraystretch{1.2}
\begin{minipage}{.48\textwidth}
     \centering
     \scalebox{.75}{
    \begin{tabular}{|l|c|c|c|}
     \hline
     & & \multicolumn{2}{c|}{Outputs}\\ \cline{3-4}
     & & \multicolumn{1}{c|}{Out\_1} & \multicolumn{1}{c|}{Out\_2}\\
     \hline
     \parbox[t]{3mm}{\multirow{4}{*}{\rotatebox[origin=c]{90}{Inputs}}} & In\_1 & A1 & A1\\
     & In\_2 & A1 & A1\\
     & In\_3 & NA2 & A1\\
     & In\_4 & NA2 & A1\\
     \hline
     \end{tabular}}
     \vspace{6pt}
     \caption{Example of a CEM.
     }
     \label{tab:CEM}
\end{minipage}
\begin{minipage}{.48\textwidth}
    \centering
    \scalebox{.9}{\begin{tabular}{|l|c|c|c|}
     \hline
     & & \multicolumn{2}{c|}{Outputs}\\ \cline{3-4}
     & & Out\_1 & Out\_2\\
     \hline
     \parbox[t]{3mm}{\multirow{3}{*}{\rotatebox[origin=c]{90}{Inputs}}} & In\_1 & Reset & Reset\\
     & In\_2 & Set & Reset\\
     & In\_3 & Set & Set\\
     \hline
     \end{tabular}}
     \vspace{5pt}
     \caption{Example of an I/O matrix.
     }
     \label{tab:IOmatrix}
\end{minipage}
\end{table}
\vspace{-60pt}
\begin{table}
\centering
\def\arraystretch{1.2}
      \centering
      \begin{minipage}{0.45\linewidth}
          \begin{figure}[H]
        \centering
    	\scalebox{.6}{\begin{tikzpicture}[->, node distance=5cm]
    
      \node [draw, shape=ellipse, align=center, minimum width=1.5cm] (mode1) {Mode\_1};  
      \node [draw, shape=ellipse, align=center, minimum width=1.5cm] (mode2) [right of=mode1, align=center] {Mode\_2};
        
      \path[every node/.style={font=\sffamily\small}]
        (mode1) edge [bend right=10] node [below, align=center] {Request\_Mode\_2} (mode2)
        (mode2) edge [bend right=10] node [above, align=center] {Request\_Mode\_1} (mode1)
    
        (mode1) edge [loop above] node [above, align=center] {$\neg$ Request\_Mode\_2} (mode1)
        (mode2) edge [loop above] node [above, align=center] {$\neg$ Request\_Mode\_1} (mode2)
        ;
        	
    \end{tikzpicture}}
    	\caption{Example of a state machine.
        }
    	\label{fig:stateMachine}
    \end{figure}
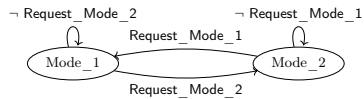
      \end{minipage}
      \hspace{0.05\linewidth}
      \begin{minipage}{0.45\linewidth}
          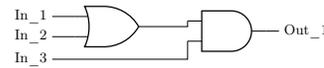
\begin{figure}[H]
            \centering
        	\scalebox{.6}{\begin{tikzpicture}
        \ctikzset{
            logic ports=ieee,
            logic ports/scale=0.8,
            logic ports/fill=white
        }

        \node[or port, number inputs=2] (ORa) at (0.5,0){};
        \node[and port, number inputs=2] (ANDa) at (3,-0.1){};
         
        \draw (ORa.out) -| (ANDa.in 1){};
        
        \draw (ANDa.out) node (AO1)   [anchor=east,xshift=1.5cm]  {Out\_1};
        \draw (ANDa.out) -- (AO1); 
        \draw (ORa.in 1) node (O1)   [anchor=east,xshift=-.5cm]  {In\_1};
        \draw (O1) -- (ORa.in 1);
        \draw (ORa.in 2) node (O2)   [anchor=east,xshift=-.5cm]  {In\_2};
        \draw (O2) -- (ORa.in 2);
        
        \draw (ORa.in 2) node (A1)   [anchor=east,xshift=-.5cm,yshift=-.5cm]  {In\_3};
        \draw (A1) -| (ANDa.in 2);
        
        \end{tikzpicture}
        }
        	\vspace{11pt}
            \caption{Example of a logic diagram.
         }
        	\label{fig:circuit}
        \end{figure}
          
      \end{minipage}
  \end{table}
\vspace{-30pt}

\begin{itemize}
    \item A~\textit{logic diagram} visually represents logical relationships.
    Figure~\ref{fig:circuit} depicts an example, representing $\mathit{Out\_1}=(In\_1\vee In\_2)\land In\_3$.
    \item An~\textit{assertion} expresses a property of a program at a particular point in the code's execution. Although this is typically used during software development, it can be used to formalize requirements. They are particularly helpful in expressing safety properties, i.e., a state can never be reached. 
    
\end{itemize}

\subsection{Formal verification}
\label{subsec:verification}

Formal verification as a service not only verifies the PLC program but also helps to amend errors in the requirements, helping requirement engineers understand the PLC program's behavior better. 
To formally verify PLC programs after the requirements are formalized, PLCverif~\cite{blancovinuela:icalepcs2019,lopez:ICALEPCS21} was used. The reasons for using it are that it is actively developed, has high coverage of PLC languages, uses state-of-the-art model checkers, has been used in real systems, and is partially automated. Other solutions, such as Arcade.PLC~\cite{arcade}, MODCHK~\cite{modchk}, PLCInspector~\cite{plcinspector}, ST\textsc{bmc}~\cite{Lee2022BoundedMC} or the ones in~\cite{modelchecking_plcs} lack some of these capabilities.

Other papers show how to use PLCverif (\cite{Fernandez:ICALEPCS2021,Adiego2015journal,adiego:ICALEPCS2017}), so due to space constraints, we will focus on the modeling of time-dependent know-how-protected functions.

PLC programming platforms such as TIA Portal~\cite{TIAportal} for Siemens PLCs include know-how-protected functions to simplify some tasks of the PLC developer. These are proprietary functions whose behavior is hidden by the manufacturer. To verify PLC programs that use these functions, it is necessary to understand their behavior precisely. 
Functions that involve time are particularly challenging, as they require the propagation of signal values across successive PLC cycles.

To model these functions, we propose a method that combines simulations and formal verification (Figure~\ref{fig:reverseEngineering}). This process produces a model in PLC code of the know-how-protected function using transparent functions and operators that can be used in PLCverif as part of the verification of the whole PLC program.

\begin{figure}[t]
    \centering
    \begin{minipage}[c]{.48\textwidth}
        \resizebox{.9\textwidth}{!}{\begin{tikzpicture}[->, node distance=1cm,connection/.style={draw,circle,fill=black,inner sep=0.8pt}]

  \node[rectangle, draw, minimum width=3.1cm, minimum height=.9cm, align=center](documentation) at (0,0){Documentation\\analysis};
  \node[rectangle, draw, minimum width=3.1cm, minimum height=.9cm, align=center](simulation) at (4,0){Function simulation};
  \node[rectangle, draw, minimum width=3.3cm, minimum height=.9cm, align=center](time) at (2,-1.05){Timing diagram\\creation};
  \node[rectangle, draw, minimum width=3.3cm, minimum height=.7cm, align=center](assertions) at (4,-2.1){Assertions and inputs\\generation};
  \node[rectangle, draw, minimum width=3.3cm, minimum height=.7cm, align=center](modeling) at (0,-2.1){Function modeling};
  \node[rectangle, draw, minimum width=3.3cm, minimum height=.7cm, align=center](verification) at (2,-3){Formal verification};

  \draw[->] (documentation.east) -- (simulation.west);
  \draw[->] (documentation.south) -- (time.95);
  \draw[->] (documentation.south) -- (modeling.north);
  \draw[->] (simulation.south) -- (time.85);
  \draw[->] (time.south) -- (assertions.north);
  \draw[->] (assertions.south) -- (verification.north);
  \draw[->] (time.south) -- (modeling.north);
  \draw[->] (modeling.south) -- (verification.north);

  \draw[-] (verification.south) --++(0,-.2) node[connection,pos=1]{};
  \draw[->] (verification.south)++(0,-.2) --++(4,0) --++(0,4.25) -| (simulation.north); 
  \draw[->] (verification.south)++(0,-.2) --++(-4,0) --++(0,4.25) -| (documentation.north); 
       
\end{tikzpicture}}
        \caption{Proposed diagram to model a know-how-protected built-in function.}
        \label{fig:reverseEngineering}
    \end{minipage}\noindent\hfill
    \begin{minipage}[c]{.48\textwidth}
        \raggedright
        \begin{tikztimingtable}[%
            timing/dslope=0.1,
            timing/.style={x=5ex,y=2ex},
            x=0ex,
            timing/rowdist=3ex,
            timing/name/.style={font=\sffamily\scriptsize}
        ]
            \busref{\#cycle}     & 1D{1} 1D{2} 1D{3} 1D{4} 1D{5} 1D{6} \\
            \busref{ON} & 1L 3H 2L \\
            \busref{FEEDBACK}    & 6H\\
            \busref{ACK}   & 3U 2L 1H \\
            \busref{ERROR}  & 3L 2H 1L \\
            \extracode
            \begin{pgfonlayer}{background}
            \begin{scope}[semitransparent ,semithick]
            \vertlines[darkgray,dotted]{0.5,1.5 ,...,5.0}
            \end{scope}
            \end{pgfonlayer}
        \end{tikztimingtable}
        \caption{Example of a timing diagram for a simplified version of the FDBACK function.}
        \label{fig:timingDiagram}
    \end{minipage}
\end{figure}

This process can be considered automata learning since the internal representation of PLCverif uses control flow automata (CFA)~\cite{Beyer2007}. In fact, the conditions of the PLC program and the assignments are translated into transitions and assignments in the CFA. However, since no tools generate PLC code from automata, learning the PLC code directly was deemed more efficient. Furthermore, having the PLC code allowed us to verify it without any extra effort with PLCverif, and to include it directly in the verification of the whole PLC project. 

To explain this process (Figure~\ref{fig:reverseEngineering}), we will use a simplified version of the know-how protected \texttt{FDBACK} function from TIA Portal~\cite[section 13.3.7]{simatic2023Safety}. It checks if the inputs \texttt{ON}=0 and \texttt{FEEDBACK}=1, and produces an error otherwise. It is used to monitor systems. 
This example is particularly relevant since it involves time, its documentation is complex, and it was often used in our case study. 

\begin{itemize}
    \item \textit{Documentation analysis}. Our simplified \texttt{FDBACK} function checks whether input \texttt{ON}=0 and input \texttt{FEEDBACK}=1. Output \texttt{ERROR} becomes 1 if this does not happen after a given maximum time (e.g., two PLC cycles). Once \texttt{ERROR}=1, an acknowledgment \texttt{ACK} is necessary to reset it.
    
    \item \textit{Function simulation}. Given the documentation, a simple PLC program is created to simulate the given function (cf.\ifextended\  Appendix~\ref{sec:appAfunctionSimulation}\else~\cite[Appendix~A.1]{lopez:NFM25_extended}\fi).

    \item \textit{Timing diagram creation}. From the simulation of the function and the documentation, we produce a timing diagram. The input variables are changed manually to capture all the possible behaviors from the documentation. Figure~\ref{fig:timingDiagram} exemplifies a timing diagram for this function. Initially, \texttt{ON}=0 and \texttt{FEEDBACK}=1, thus \texttt{ERROR}=0. Then \texttt{ON} turns 1, leading to \texttt{ERROR}=1 after the maximum time (two cycles) is reached. Although \texttt{ON} becomes 0 again, \texttt{ERROR} keeps its value until there is an acknowledgment (\texttt{ACK}=1) (cf.\ifextended\  Appendix~\ref{sec:appAtimingDiagram}\else~\cite[Appendix~A.2]{lopez:NFM25_extended}\fi). 
    
    \item \textit{Assertions and inputs generation}. The timing diagram is automatically translated for every cycle into assignments for the inputs and assertions for the outputs. Since \texttt{ACK} is non-deterministic in the first three cycles, no value is assigned to it. For the first cycle, the assignments are $\mathit{FEEDBACK}:=1,\mathit{ON}:=0$, and the assertion is $\mathit{cycle}=1 \to \neg \mathit{ERROR}$ (cf.\ifextended \ Appendix~\ref{sec:appAassertionsGeneration}\else~\cite[Appendix~A.3]{lopez:NFM25_extended}\fi).

    \item \textit{Function modeling}. Given the documentation, a simple PLC program is created to simulate the given function (cf.\ifextended\ Appendix~\ref{sec:appAfunctionModeling}\else~\cite[Appendix~A.4]{lopez:NFM25_extended}\fi).

    \item \textit{Formal verification}. The modeled function is verified with respect to the generated assertions. This ensures that the modeled function behaves as the one in TIA Portal with respect to all the simulated scenarios (cf.\ifextended\  Appendix~\ref{sec:appAformalVerification}\else~\cite[Appendix~A.5]{lopez:NFM25_extended}\fi).

\end{itemize}

This process continues iteratively until no discrepancies are found between the documentation, the timing diagrams, and the PLC model. The modeling of the original \texttt{FDBACK} function resulted in about 100 lines of code~\cite[Line~843]{builtin-plcverif}.

\section{Case study}
\label{sec:caseStudy}

The approach presented in this paper was applied to verify the Personnel Access System (PAS) of the  FAIR accelerator facility at the \textit{GSI Helmholtzzentrum für Schwerionenforschung}~\cite{gsi}. PAS~\cite{salinas2023ipac} is a very critical system that prevents personnel from entering areas exposed to particle beams and their radiation. Thus, a failure in the PAS PLC program could have very severe consequences.
This PLC program is highly configurable, making exhaustive testing unfeasible due to the enormous number of combinations in the PLC program. Also, it is developed using TIA Portal, hence, it utilizes know-how-protected functions.

Due to CERN's expertise in the verification of different PLC projects~\cite{Fernandez:ICALEPCS2021,adiego:ICALEPCS2017} and the continuous development of PLCverif, GSI trusted CERN to verify its PAS PLC project. The collaboration was set up as described in Section~\ref{sec:approach} with three independent teams:
\begin{inlinelist}
    \item Requirement engineers (GSI).
    \item Formal verification engineers (CERN).
    \item Developers (a different team at CERN).
\end{inlinelist}

A summary of the results produced by this collaboration is shown below:
\begin{itemize}
    \item The requirements were formalized according to Section~\ref{subsec:specification}, leading to a better understanding of the desired program behavior and less ambiguities;
    \item The PLC program was fully aligned with the formal requirements, amending detected discrepancies (cf.\ifextended\ Appendix~\ref{sec:appDiscrepancies}\else~\cite[Appendix~B]{lopez:NFM25_extended}\fi\ for examples of found discrepancies);
    \item PLCverif was enhanced to support additional know-how-protected functions, including \texttt{FDBACK}, \texttt{CTUD}, \texttt{ESTOP1}, and \texttt{FDB\_TIME}~\cite{simatic2023Safety}.
    They are now included in the set of covered functions by PLCverif~\cite{plcverifweb} (delivered together with PLCverif in the \texttt{builtin.scl} file) and can be used in future projects.
\end{itemize}

\section{Conclusion}
\label{sec:conclusion}

The presented approach for formal verification as a service can help to detect errors early, reduce ambiguity, and improve requirements precision.
To the best of our knowledge, this is the first time an organization trusted another to formally verify a complete, real-world, safety-critical PLC project (other collaborations like ITER-CERN \cite{adiego:ICALEPCS2017} focused on the verification of specific modules). We hope the presented approach demonstrates that formal methods are feasible, beneficial, and compliant with functional safety standards in safety-critical PLC projects, enabled by organizational collaborations.

As part of our future work, we will seek a more automated process of modeling know-how-protected functions to increase the coverage of PLCverif. We will also work on the automation between requirement specification and verification, ultimately leading to correct-by-construction code generation. This is not a straightforward path due to different challenges, such as a lack of formal tools, legacy systems, established workshops, regulations, and the need for training.

\section*{Acknowledgments}
\hfill
\setlength{\intextsep}{0pt}
\begin{wrapfigure}{l}{.8cm}
\includegraphics[width=1cm]{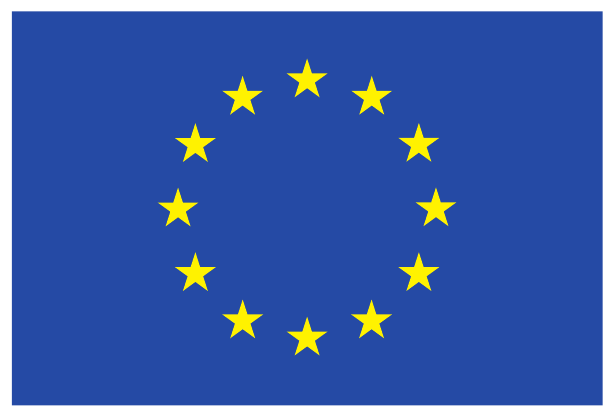}
\end{wrapfigure}
\noindent The project leading to this application has received funding from the European Union's Horizon 2020 research and innovation programme under grant agreement No 101034440 and by the \href{https://taiger.logic.at}{WWTF project ICT22-023}.

\bibliographystyle{splncs04}

\ifextended
\newpage
\noindent

\section*{Appendix}
\appendix

\section{Modelling of the \texttt{FDBACK} function}
\label{sec:appendixFDBACK}

In this appendix, we will show in detail an example of modeling a \textit{know-how protected} function. We will use the same example as in the main text, i.e., the \texttt{FDBACK} function. For the steps of the process where we need to interact with TIA portal (function simulation and timing diagram creation), we will use the original \texttt{FDBACK} function. For the other parts, we will use the simplified version that we presented in the section~\ref{subsec:verification} to simplify the explanation.

\subsection{Function simulation}
\label{sec:appAfunctionSimulation}

We created a simple TIA Portal project containing only the function we want to model. Figure~\ref{fig:project} shows the small project structure that was used to simulate the original \texttt{FDBACK} function. The inputs and outputs of the \texttt{FDBACK} function (called by the \texttt{Main\_Safety\_RTG1} Function Block) are shown in Figure~\ref{fig:tiaPortalProjectSimulation} and can be enabled for simulation.
\vspace{6pt}
\begin{figure}
    \centering
    \begin{minipage}[c]{.31\textwidth}
        \includegraphics[width=.98\linewidth]{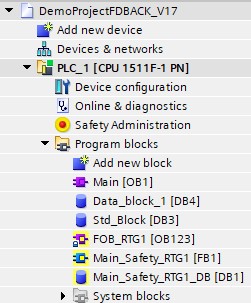}
         \caption{Project structure in TIA Portal to simulate \texttt{FDBACK} function.}
    \label{fig:project}
    \end{minipage}\noindent\hfill
    \begin{minipage}[c]{.67\textwidth}
    \centering
    \includegraphics[width=.77\linewidth]{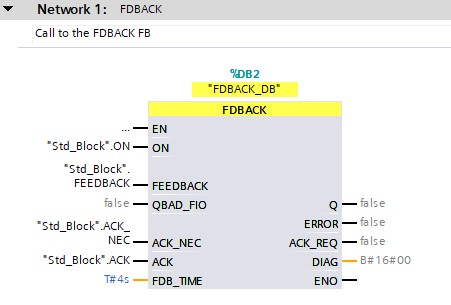}
    \caption{\texttt{FDBACK} function interface call in the TIA Portal program.}
    \label{fig:tiaPortalProjectSimulation}
    \end{minipage}
\end{figure}

\subsection{Timing diagram creation}
\label{sec:appAtimingDiagram}

For this project, we used the PLCSIM simulator~\cite{PLCSIM} provided by Siemens, which is integrated into TIA portal. 
To simulate the \texttt{FDBACK} function, we manually forced its inputs according to what we wanted to check from the documentation. Once the inputs were set, the outputs were observed. Note that safety inputs (in yellow in \Cref{fig:tiaPortalInputOutputVarsSimulation2,fig:tiaPortalInputOutputVarsSimulation3,fig:tiaPortalInputOutputVarsSimulation4}) cannot be forced on the simulator. For this reason, we used a standard Data Block (DB) called \texttt{std\_Block}, and we assigned the variables of this DB to the safety inputs of the \texttt{FDBACK} function. By forcing the \texttt{std\_Block} variables, we can change the input values of the \texttt{FDBACK} function. 

 
Since PLCSIM does not provide a timing diagram as such, from the manual simulations, a timing diagram was created manually (cf. table from Figure~\ref{fig:spreadsheetFDBACK}). Eventually, this could be automated by using the TIA Openness API~\cite{Openness} and the PLCSIM advanced simulator~\cite{PLCSIMadvanced}, which is only available for the S7-1500 PLC series. The latter provides continuous, cycle-by-cycle, and time-synchronized execution modes and comes with a C\# API, which can be used to automate the simulation and execution process~\cite{blanco:icalepcs19-wepha018}.
Siemens also provides \texttt{OpennessScripter}~\cite{OpennessScripter}, which is a tool to simplify the use of the TIA Portal Openness interface.

Several simulations were performed following all the situations that the documentation describes. This is an iterative process; more simulations were included when some behaviors were unclear. 

We will now show three simulation scenarios that can complete a simple timing diagram. We set values for the input variables of the \texttt{FDBACK} function and observe its outputs. These simulations show how the system goes through the following states: \begin{inlinelist}
    \item no error,
    \item error,
    \item error acknowledged $\rightarrow$ error removed.
\end{inlinelist}
 We will only focus on the variables that are part of our simplified \texttt{FDBACK} function. That is, inputs=\{\texttt{ON},\texttt{FEEDBACK},\texttt{ACK}\}, and outputs=\{\texttt{ERROR}\}.

\begin{enumerate}
    \item Figure~\ref{fig:tiaPortalInputOutputVarsSimulation2}. There is no error since \texttt{ON}=0 and \texttt{FEEDBACK}=1.
    \begin{multicols}{2}
        \begin{itemize}
            \item \texttt{ON} $\leftarrow$ \texttt{FALSE}
            \item \texttt{FEEDBACK} $\leftarrow$ \texttt{TRUE}
            \item \texttt{ACK} $\leftarrow$ \texttt{FALSE}
        \end{itemize}
        \begin{itemize}
            \item \texttt{ERROR} $=$ \texttt{FALSE}
        \end{itemize}
    \end{multicols}
\end{enumerate}

\begin{figure}[H]
    \centering
    \includegraphics[width=.85\linewidth]{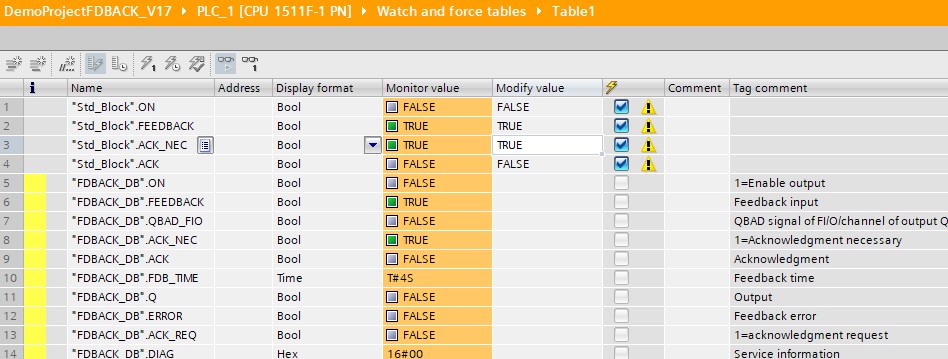}
    \caption{Input and output variables of the original \texttt{FDBACK} function in TIA Portal}
    \label{fig:tiaPortalInputOutputVarsSimulation2}
\end{figure}
\vspace{6pt}
\begin{figure}[H]
    \centering
    \includegraphics[width=.85\linewidth]{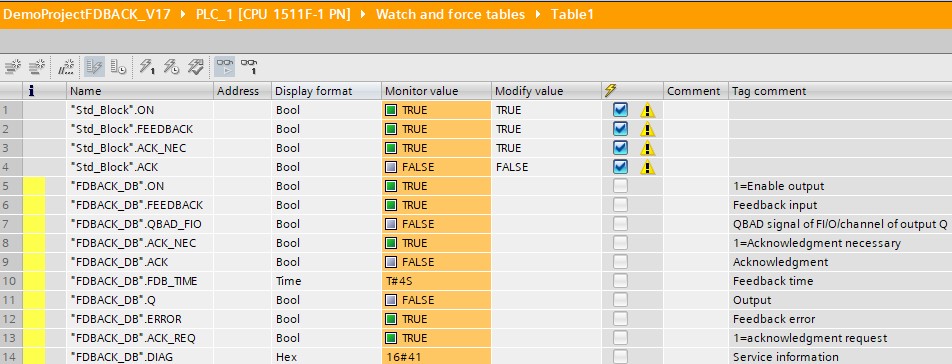}
    \caption{Input and output variables of the original \texttt{FDBACK} function in TIA Portal}
    \label{fig:tiaPortalInputOutputVarsSimulation3}
\end{figure}

\begin{enumerate}\setcounter{enumi}{1}
    \item Figure~\ref{fig:tiaPortalInputOutputVarsSimulation3}. There is an error since \texttt{ON}=1 and \texttt{FEEDBACK}=1. The screenshot has been taken after waiting for two PLC cycles with those inputs so that the \texttt{ERROR} becomes 1.
    \begin{multicols}{2}
        \begin{itemize}
            \item \texttt{ON} $\leftarrow$ \texttt{TRUE}
            \item \texttt{FEEDBACK} $\leftarrow$ \texttt{TRUE}
            \item \texttt{ACK} $\leftarrow$ \texttt{FALSE}
        \end{itemize}
        \begin{itemize}
            \item \texttt{ERROR} $=$ \texttt{TRUE}
        \end{itemize}
    \end{multicols}
\end{enumerate}

\begin{enumerate}\setcounter{enumi}{2}
    \item Figure~\ref{fig:tiaPortalInputOutputVarsSimulation4}. There is no error anymore because now \texttt{FEEDBACK}=\texttt{TRUE} and \texttt{ON}=\texttt{FALSE}, and the error has been acknowledged with a rising edge of \texttt{ACK}.
    \begin{multicols}{2}
    \begin{itemize}
        \item \texttt{ON} $\leftarrow$ \texttt{FALSE}
        \item \texttt{FEEDBACK} $\leftarrow$ \texttt{TRUE}
        \item \texttt{ACK} $\leftarrow$ \texttt{TRUE}
    \end{itemize}
    \begin{itemize}
        \item \texttt{ERROR} $=$ \texttt{FALSE}
    \end{itemize}
    \end{multicols}
\end{enumerate}

\begin{figure}[H]
    \centering
    \includegraphics[width=.85\linewidth]{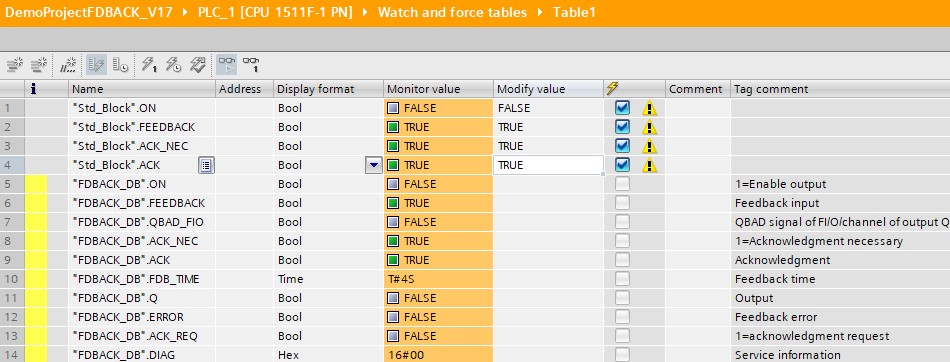}
    \caption{Input and output variables of the original \texttt{FDBACK} function in TIA Portal}
    \label{fig:tiaPortalInputOutputVarsSimulation4}
\end{figure}
\vspace{6pt}
These consecutive simulations are annotated in a table like the one from Figure~\ref{fig:spreadsheetFDBACK}, producing a timing diagram as also shown in Figure~\ref{fig:spreadsheetFDBACK}.
The generation of timing diagrams is concluded when all the aspects from the documentation are covered and no assertion fails (cf. Appendix~\ref{sec:appAassertionsGeneration} and Appendix~\ref{sec:appAformalVerification}).

\subsection{Assertions and inputs generation}
\label{sec:appAassertionsGeneration}

Once a timing diagram was created, the PLC code for the verification of the model of a know-how-protected function was generated automatically. It contains the statements to set the input variables to the corresponding values and the assertions to check the outputs. 

The spreadsheet used for the simplified \texttt{FDBACK} function is shown in Figure~\ref{fig:spreadsheetFDBACK}. The table corresponds to the encoding of the timing diagram from Figure~\ref{fig:timingDiagram}. The two blocks of code correspond to the assignments to the input variables and to the assertions. One can see that the assertions do not contain any input variable since they are already set to the correct value with the assignment statements. If there is no assignment for an input variable, it can take any value non-deteministically.

\begin{figure}[H]
    \centering
    \includegraphics[width=0.9\linewidth]{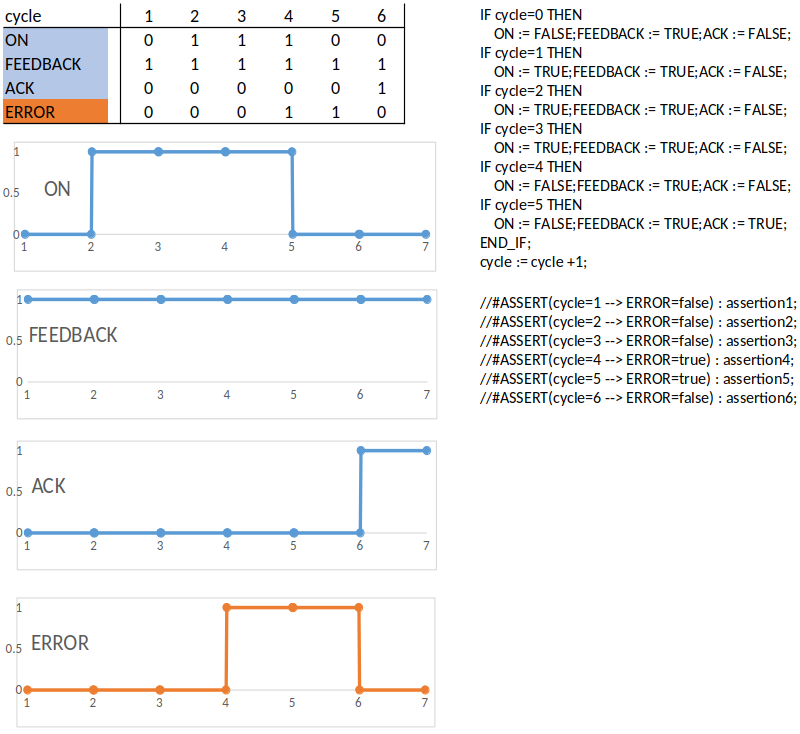}
    \caption{Screenshot of the spreadsheet to generate the code to verify the model of the simplified know-how protected \texttt{FEEDBACK} function.}
    \label{fig:spreadsheetFDBACK}
\end{figure}
\vspace{6pt}
The generated code is the one used in Listing~\ref{codeVerificationFDBACK} to verify the function. 
For the modeling of the original \texttt{FDBACK} function, six spreadsheets with 25 cycles each were used.

Although it could be possible to have a unique assertion as the one shown below for each timing diagram, it is preferable to split it into smaller assertions, as in the PLC code from Listing~\ref{codeVerificationFDBACK}, to be able to find the root cause of discrepancies faster.

\begin{equation*}
\begin{split}
& (\mathit{cycle}=1 \to \neg \mathit{ERROR}) \land (\mathit{cycle}=2 \to \neg \mathit{ERROR}) \land (\mathit{cycle}=3 \to \neg \mathit{ERROR}) \land\\
& (\mathit{cycle}=4 \to \mathit{ERROR}) \land (\mathit{cycle}=5 \to \mathit{ERROR}) \land (\mathit{cycle}=6 \to \neg\mathit{ERROR}) 
\end{split}
\end{equation*}

\subsection{Function modeling}
\label{sec:appAfunctionModeling}

The model in PLC code of the simplified know-how-protected \texttt{FDBACK} function can be seen in Listing~\ref{codeFDBACKscl}. This model was done manually according to the documentation and simulations. One can see that even though the requirement looks simple, its implementation is not trivial due to the timing aspect. Furthermore, this is just a simplified version of the real one, whose model in PLC code was implemented in about 100 lines of code~\cite[Line~843]{builtin-plcverif}.

\vspace{.2cm}
\begin{lstlisting}[caption={PLC code modeling the simplified \textit{know-how protected} \texttt{FDBACK} function.}, style=assertion, label=codeFDBACKscl]
FUNCTION_BLOCK FDBACK_simplified
    VAR_INPUT
        ON : BOOL;
            FEEDBACK : BOOL;
        ACK : BOOL;
    END_VAR
    VAR
    // Elapsed Time (ET) variable is a Timer On Delay. 
    //  It sets ET.Q to true after a given time
        ET : TON;
    END_VAR

    VAR_OUTPUT
        ERROR : BOOL := FALSE;
    END_VAR

BEGIN
    IF ERROR THEN // manual acknowledgement
        IF ACK THEN
            ET(IN := FALSE); // reset timer
            ERROR := FALSE; // reset the error
        END_IF;
    ELSIF NOT (NOT ON AND FEEDBACK) THEN
        // start timer (ET is the Elapsed Time function)
        ET(IN := TRUE, PT := 2*200);// 2 cycles (each safety cycle is 200ms)
        IF ET.Q THEN // if waiting time is over
            ERROR := TRUE;
        END_IF;
    ELSE
        ET(IN := FALSE); // reset timer
    END_IF;

END_FUNCTION_BLOCK
\end{lstlisting}
\vspace{6pt}
\begin{figure}[H]
    \centering
    \scalebox{.68}{\begin{tikzpicture}[->, node distance=6cm]

  \node [state, accepting, align=center, minimum width=1.5cm] (s0) {$\neg\mathit{ERROR}$};
  \node (init) [above left=1cm and 1cm of s0] {};
  \node [state, align=center, minimum width=1.5cm] (s1) [right of=s0, align=center] {$\neg\mathit{ERROR}$};
  \node [state, accepting, align=center, minimum width=1.5cm] (s2) [right of=s1, align=center] {$\mathit{ERROR}$};
    
  \path[every node/.style={font=\sffamily\small}]
    (init) edge (s0)
  
    (s0) edge [bend right=10] node [below, align=center] {$\neg(\neg\mathit{ON} \land \mathit{FEEDBACK})$} (s1)
    
    (s1) edge [bend right=10] node [above, align=center] {$\neg\mathit{ON} \land \mathit{FEEDBACK}$,\\$t:=0$} (s0)
    
    (s1) edge node [above, align=center] {$\neg(\neg\mathit{ON} \land \mathit{FEEDBACK})$,\\$t>2$} (s2)
    
    (s2) edge [loop above] node [above, align=center] {$\neg\mathit{ACK}$} (s2)
    
    (s0) edge [loop above] node [above, align=center] {$\neg\mathit{ON} \land \mathit{FEEDBACK}$,\\$t:=0$} (s0)

    (s1) edge [loop above] node [above, align=center] {$\neg\mathit{ON} \land \mathit{FEEDBACK}$, $t\leq 2$} (s1)
    
    (s2) edge [bend left=30] node [above, align=center] {$\mathit{ACK}$} (s0)
    
    ;
        
\end{tikzpicture}}
    \caption{Timed automaton representing the simplified \texttt{FDBACK} function.}
    \label{fig:automatonFDBACK}
\end{figure}
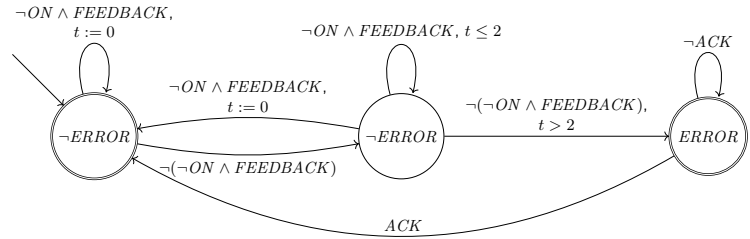

As mentioned in section~\ref{subsec:verification}, the problem of modeling know-how-protected functions can be understood in terms of automata learning.
For the simplified \texttt{FDBACK} function, the corresponding timed automaton that can be extracted from the PLC code is shown in Figure~\ref{fig:automatonFDBACK}.

\subsection{Formal verification}
\label{sec:appAformalVerification}

In order to verify that the model is working as expected, the PLC code shown in Listing~\ref{codeVerificationFDBACK} was used. It basically consists of a set of consecutive calls to the model of the function \texttt{FDBACK\_simplified."FDBACK\_simplified\_inst"()}. Each call represents a PLC cycle. For each cycle, the input variables are set to the values according to the timing diagram (see Figure~\ref{fig:timingDiagram}). If a variable is not set, its value is non-deterministic, as with \texttt{ACK}. At the end of each cycle, it is checked if the output variable \texttt{ERROR} has the same value as in the timing diagram. 
\vspace{6pt}
\begin{lstlisting}[caption={PLC code used to verify the simplified \textit{know-how protected} \texttt{FDBACK} function.}, style=assertion, label=codeVerificationFDBACK]
DATA_BLOCK "FDBACK_simplified_inst" FDBACK_simplified
BEGIN
END_DATA_BLOCK

FUNCTION_BLOCK call_FDBACK_simplified
	VAR
           cycle : INT := 1;
	END_VAR
BEGIN

// setting the input variables according to the timing diagram
// in the cycles where ACK is not set, its value is non-deterministic
IF cycle=1 THEN 
    "FDBACK_simplified_inst".FEEDBACK:= TRUE;  
    "FDBACK_simplified_inst".ON := FALSE;
ELSIF cycle=2 THEN 
    "FDBACK_simplified_inst".FEEDBACK := TRUE;  
    "FDBACK_simplified_inst".ON := TRUE;
ELSIF cycle=3 THEN 
    "FDBACK_simplified_inst".FEEDBACK := TRUE;  
    "FDBACK_simplified_inst".ON := TRUE;
ELSIF cycle=4 THEN 
    "FDBACK_simplified_inst".FEEDBACK := TRUE;  
    "FDBACK_simplified_inst".ON := TRUE;  
    "FDBACK_simplified_inst".ACK := FALSE;
ELSIF cycle=5 THEN 
    "FDBACK_simplified_inst".FEEDBACK := TRUE; 
    "FDBACK_simplified_inst".ON := FALSE;  
    "FDBACK_simplified_inst".ACK := FALSE;
ELSIF cycle=6 THEN 
    "FDBACK_simplified_inst".FEEDBACK := TRUE; 
    "FDBACK_simplified_inst".ON := FALSE;  
    "FDBACK_simplified_inst".ACK := TRUE;
END_IF;

// check with assertions that the output variables have the same
// values than in the timing diagram
 
FDBACK_simplified."FDBACK_simplified_inst"() ;

//#ASSERT(cycle=1 --> ("FDBACK_simplified_inst".ERROR = FALSE)) : assertion1; 
//#ASSERT(cycle=2 --> ("FDBACK_simplified_inst".ERROR = FALSE)) : assertion2; 
//#ASSERT(cycle=3 --> ("FDBACK_simplified_inst".ERROR = FALSE)) : assertion3; 
//#ASSERT(cycle=4 --> ("FDBACK_simplified_inst".ERROR = TRUE)) : assertion4; 
//#ASSERT(cycle=5 --> ("FDBACK_simplified_inst".ERROR = TRUE)) : assertion5; 
//#ASSERT(cycle=6 --> ("FDBACK_simplified_inst".ERROR = FALSE)) : assertion6; 
 
cycle := cycle +1;
 
END_FUNCTION_BLOCK
\end{lstlisting}

As already mentioned, PLCverif was used to verify the PLC code modeling know-how protected functions. This process was straightforward once the code was generated. However, it is essential to highlight how the time was treated with PLCverif. Since the time of the PLC cycle is not relevant for verification purposes -- what is important is when the output of the timer is activated -- a fixed time of the PLC cycle was fixed. In this case, we used the usual safety time of the PLC cycle of $\mathit{T\_CYCLE}=100$ms. This can be seen in Figure~\ref{fig:t-cycle} from the verification case of PLCverif. 

Furthermore, it is important to allow as many cycles as needed to be able to trigger the output of the timer. That is, if the output is triggered after $T$ milliseconds of being started, then we should have at least $c_{c}=\mathit{int}(T/\mathit{T\_CYCLE})+1$ cycles. We should have also at least the number of cycles that we have in the timing diagram $c_t$. Thus, the number of loop unwinding for a bounded model checker like CBMC should be $\max \{c_c,c_t\}$. In our case, $c_c=2$ ($200$ms$/100$ms) and $c_t=6$, thus we set it to 6. This can be seen in Figure~\ref{fig:loop-unwinding} from the verification case of PLCverif.

Once the verification case is executed and we get that all assertions are satisfied, as shown in Figure~\ref{fig:verReport}, the modeling process is finished. If PLCverif reports a violation of an assertion, it will also give us a counterexample. Then, we need to investigate where the error is coming from and amend it so that the model aligns with the timing diagram.

\begin{figure}[H]
    \centering
    \includegraphics[width=.75\linewidth]{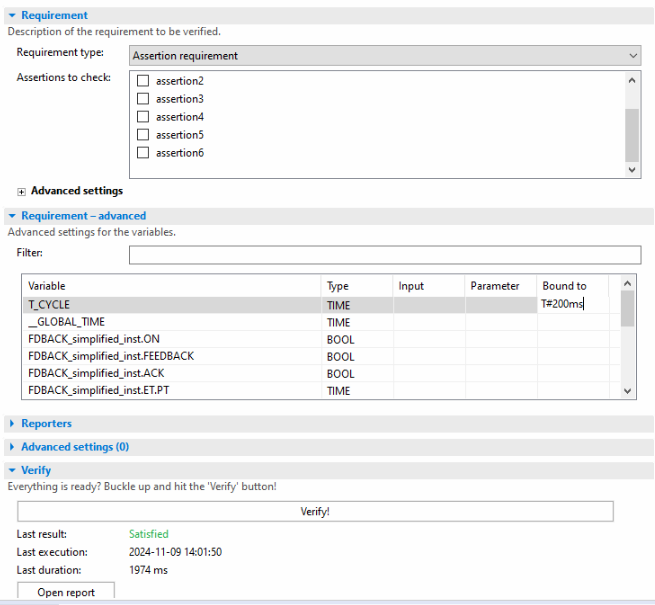}
    \caption{Verification case of PLCverif. The time of the cycle is set to $100$ms.}
    \label{fig:t-cycle}
\end{figure}
\vspace{6pt}
\begin{figure}[H]
    \centering
    \includegraphics[width=.75\linewidth]{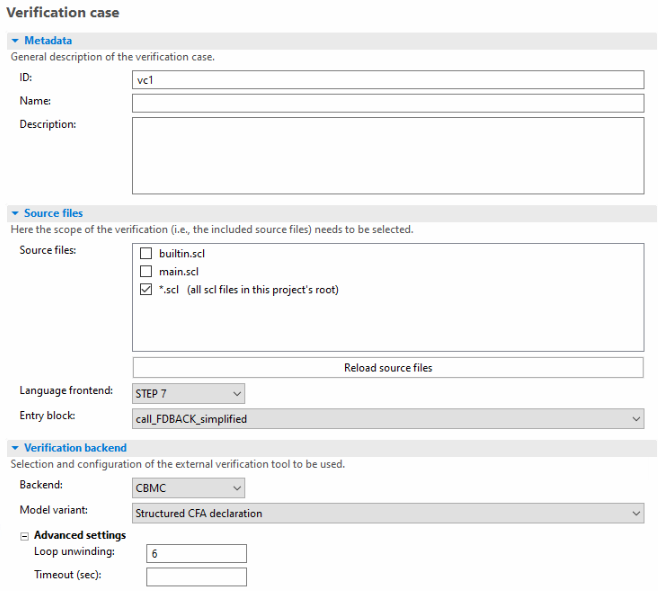}
    \caption{Verification case of PLCverif. The loop unwinding is set to 6 to cover all the cycles in the timing diagram.}
    \label{fig:loop-unwinding}
\end{figure}

\begin{figure}[H]
    \centering
    \includegraphics[width=0.75\linewidth]{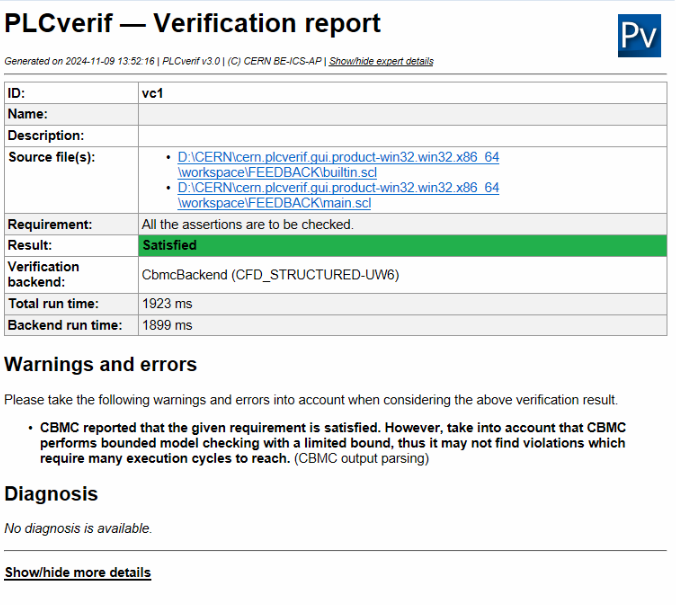}
    \caption{Verification report of PLCverif. All the assertions are satisfied.}
    \label{fig:verReport}
\end{figure}

\section{Examples of found discrepancies}
\label{sec:appDiscrepancies}

This section summarizes the discrepancies found during the verification of different PLC projects by CERN, not only during the verification of the GSI project. The examples are simplified to show where the problem lies more easily. Most of the discrepancies can be classified into the following three buckets:

\begin{enumerate}
    \item \textit{Incomplete requirements}. This is the most common type of discrepancy found. The implementation works as expected by the requirement engineers, but the requirements have not been formalized correctly.
    \item \textit{Bugs in the PLC program}. It happens when the requirement is correct, but the implementation has an error.
    \item \textit{Minor documentation errors}. These are simple errors that are easily fixed. A reader can understand the requirements without any issues. For example, a misspelled variable would be part of this type of error.
\end{enumerate}

In the next subsection~\ref{sec:incompleteRequirements}, we will give some examples of incomplete requirements. However, we will not extend the other two types of discrepancies since they are self-explanatory.

Other types of problems can also be found during the formal verification of a PLC project. A recurrent one that appears in the early stages of the collaboration is \textit{what} to specify and \textit{how} to do it formally. Furthermore, simple things like the exact software used are sometimes not specified, as well as what happens if there are \textit{hardware failures}.

\subsection{Incomplete requirements}
\label{sec:incompleteRequirements}

The examples shown in this subsection are not exhaustive but include the majority of the most important discrepancies found. We will cover situations related to \begin{inlinelist}
    \item priorities,
    \item incomplete diagrams and tables,
    \item and lack of explanations.
\end{inlinelist} 
We will show incomplete requirements and propose solutions to complete them.

\subsubsection{Priorities.} Although it is not ideal, different requirements often express conditions for the same output variables. If no priorities are set, this can lead to ambiguities. As an example, let us take the following requirements:

\begin{minipage}{.01\textwidth}
\end{minipage}
\begin{minipage}{.99\textwidth}\noindent\hfill
\begin{enumerate}[label=(\subscript{R}{{\arabic*}})]
    \item If \texttt{v1\_up} $\to$ set \texttt{v\_out}.
    \item If \texttt{v1\_down} $\to$ reset \texttt{v\_out} 
\end{enumerate}
\end{minipage}
\vspace{.2cm}

Listing~\ref{codePriorities} shows an example of how this requirement can be implemented. In this case, $R_2$ has a higher priority than $R_1$ since if \texttt{v1\_down} is true, then \texttt{v\_out} will be true no matter the value of \texttt{v1\_up}. What is executed later has a higher priority. However, another implementation could change the order of the \texttt{IF} statements, leading to $R_1$ having a higher priority than $R_1$. This ambiguity can be solved by 
\begin{enumerate}
    \item telling which of the two requirements has a higher priority,
    \item if \texttt{v1\_up} and \texttt{v1\_down} cannot be true simultaneously (e.g., physical constraints), stating this fact,
    \item adding all the necessary variables in each requirement:
    \vspace{.2cm}
    \begin{enumerate}[label=(\subscript{R'}{{\arabic*}})]
        \item If \texttt{v1\_up} and not \texttt{v1\_down} $\to$ set \texttt{v\_out}.
        \item If \texttt{v1\_down} and not \texttt{v1\_up} $\to$ reset \texttt{v\_out}. 
    \end{enumerate}
\end{enumerate}

\begin{lstlisting}[caption={PLC code implementing a solution for ambiguous requirements where the priorities are not set.}, style=assertion, label=codePriorities]
FUNCTION_BLOCK req_priorities
    VAR_INPUT
        v1_up : BOOL;
        v1_down : BOOL;
    END_VAR
    VAR_OUTPUT
        out : BOOL;
    END_VAR
BEGIN
    IF v1_up THEN
        v_out := TRUE;
    END_IF;
    IF v1_down THEN
        v_out := FALSE;
    END_IF;
END_FUNCTION_BLOCK

\end{lstlisting}










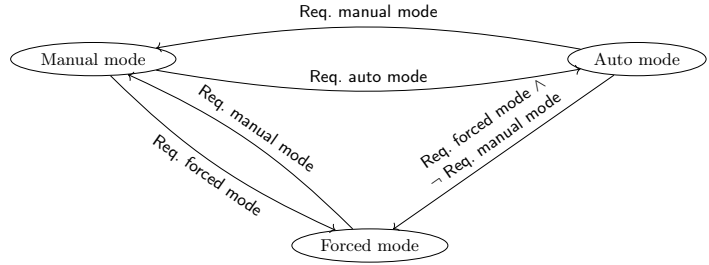
\begin{figure}[H]
    \centering
    \scalebox{.72}{\begin{tikzpicture}[->]

  \node [draw, shape=ellipse, align=center, minimum width=1.5cm] (s0) {Manual mode};
  \node [draw, shape=ellipse, align=center, minimum width=1.5cm] (s1) [below right=3cm and 3cm of s0] {Forced mode};
  \node [draw, shape=ellipse, align=center, minimum width=1.5cm] (s2) [above right=3cm and 3cm of s1, align=center] {Auto mode};
    
  \path[every node/.style={font=\sffamily\small}]
    (s0) edge [bend right=10] node [below, align=center, rotate=-35] {Req. forced mode} (s1)
    (s1) edge [bend right=10] node [above, align=center, rotate=-35] {Req. manual mode} (s0)    
    (s2) edge node [above, align=center, rotate=35] {Req. forced mode $\land$\\$\neg$ Req. manual mode} (s1)
    (s2) edge [bend right=10] node [above, align=center] {Req. manual mode} (s0)
    (s0) edge [bend right=10] node [above, align=center] {Req. auto mode} (s2)
    ;
        
\end{tikzpicture}}
    \caption{Example of an incomplete state machine.}
    \label{fig:incompleteStateMachine}
\end{figure}

\subsubsection{Incomplete diagram.}

Let us take the state machine from Figure~\ref{fig:incompleteStateMachine}. It represents how a system can change its operation mode by requesting it. From \texttt{manual mode}, it is possible to transition to \texttt{auto mode} and to \texttt{forced mode}. However, it is not specified what happens when the corresponding requests to transition to \texttt{auto mode} and to \texttt{forced mode} are both true simultaneously. This is also the case for the transitions between \texttt{auto mode} to \texttt{manual mode} and \texttt{forced mode}.
\vspace{12pt}
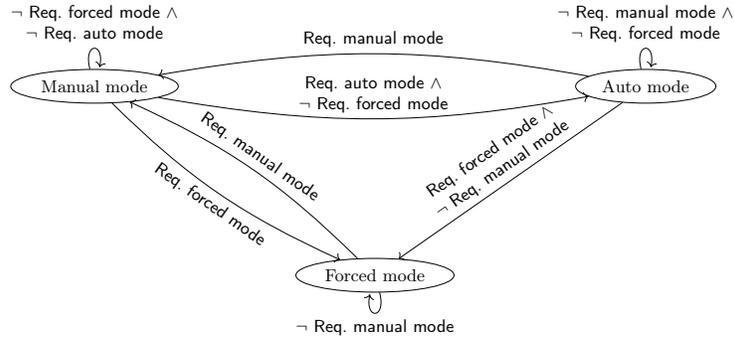
\begin{figure}[H]
    \centering
    \scalebox{.73}{\begin{tikzpicture}[->]

  \node [draw, shape=ellipse, align=center, minimum width=1.5cm] (s0) {Manual mode};
  \node [draw, shape=ellipse, align=center, minimum width=1.5cm] (s1) [below right=3cm and 3cm of s0] {Forced mode};
  \node [draw, shape=ellipse, align=center, minimum width=1.5cm] (s2) [above right=3cm and 3cm of s1, align=center] {Auto mode};
    
  \path[every node/.style={font=\sffamily\small}]
    (s0) edge [bend right=10] node [below, align=center, rotate=-35] {Req. forced mode} (s1)
    (s1) edge [bend right=10] node [above, align=center, rotate=-35] {Req. manual mode} (s0)    
    (s2) edge node [above, align=center, rotate=35] {Req. forced mode $\land$\\$\neg$ Req. manual mode} (s1)
    (s2) edge [bend right=10] node [above, align=center] {Req. manual mode} (s0)
    (s0) edge [bend right=10] node [above, align=center] {Req. auto mode $\land$\\$\neg$ Req. forced mode} (s2)
    (s0) edge [loop above] node [above, align=center] {$\neg$ Req. forced mode $\land$\\$\neg$ Req. auto mode} (s0)
    (s1) edge [loop below] node [below, align=center] {$\neg$ Req. manual mode} (s1)
    (s2) edge [loop above] node [above, align=center] {$\neg$ Req. manual mode $\land$\\$\neg$ Req. forced mode} (s2)
    ;
        
\end{tikzpicture}}
    \caption{Example of how the incomplete state machine from Figure~\ref{fig:incompleteStateMachine} can be fixed.}
    \label{fig:completeStateMachine}
\end{figure}
\vspace{6pt}
In order to fix this situation, one needs to specify all the necessary conditions for each guard so that only one transition is activated at a time. Figure~\ref{fig:completeStateMachine} shows a possible fix to the previous ambiguous state machine. Now, if the system is in \texttt{manual mode} and there is a simultaneous request to transition both to \texttt{auto mode} and to \texttt{forced mode}, the system will transition to \texttt{forced mode}.

\subsubsection{Lack of explanation.}

In some cases, we experienced situations where there was a lack of explanation about certain requirements.
\begin{itemize}
    \item \textbf{Global/Shared/Input-Output variables}. When a project comprises different modules, some variables flow from one module to another. They are part of the output variables in one module and of the inputs in other modules. Since the requirements are usually formalized per module, these variables can be treated as inputs in some modules. However, they are not free inputs for those modules in the sense that they can only take a limited set of possible values given by the output of the other module. This fact is usually not stated, leading to the violation of properties with values for those variables that are not possible. A possible way to formally verify these situations is by using assume-guarantees (possible with PLCverif) or contracts.
    
    Figure~\ref{fig:modulesVariables} shows an example in which the variable $v_1$ is an output of module 1 and an input of modules 2 and 3. In this case, $v_1$ cannot take any value and is limited to the possible values produced by module 1. Therefore, if a requirement for module 2 or 3 includes this variable, it might be violated with a value for $v_1$ that can never happen. Requirement engineers might have already in mind that the value for that variable is limited to a specific range given by module 1 but might not have specified it when writing the requirements for modules 2 and 3.

\vspace{6pt}

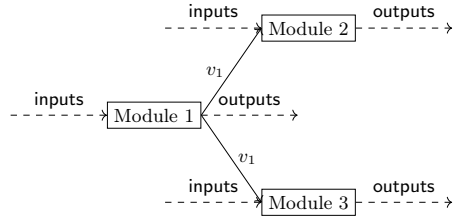
\begin{figure}[H]
    \centering
    \scalebox{.8}{\begin{tikzpicture}[->, node distance=2.5cm]

  \node [draw, shape=rectangle, align=center, minimum width=1.5cm] (s0) {Module 1};
  \node (inputss0) [left of=s0] {};
  \node (outputss0) [right of=s0] {};
  \node [draw, shape=rectangle, align=center, minimum width=1.5cm] (s1) [above right=1cm and 1cm of s0] {Module 2};
  \node (inputss1) [left of=s1] {};
  \node (outputss1) [right of=s1] {};
  \node [draw, shape=rectangle, align=center, minimum width=1.5cm] (s2) [below right=1cm and 1cm of s0, align=center] {Module 3};
  \node (inputss2) [left of=s2] {};
  \node (outputss2) [right of=s2] {};
    
  \path[every node/.style={font=\sffamily\small}]
    (s0.east) edge node [left, align=center] {$v_1$} (s1.west) 
    (s0.east) edge node [right, align=center] {$v_1$} (s2.west)
    (inputss0) edge [dashed] node [above, align=center] {inputs} (s0)
    (s0) edge [dashed] node [above, align=center] {outputs} (outputss0)
    (inputss1) edge [dashed] node [above, align=center] {inputs} (s1)
    (s1) edge [dashed] node [above, align=center] {outputs} (outputss1)
    (inputss2) edge [dashed] node [above, align=center] {inputs} (s2)
    (s2) edge [dashed] node [above, align=center] {outputs} (outputss2)
    ;
        
\end{tikzpicture}}
    \caption{Example of modules in which a variable $v_1$ is the output from one of them and the input for the other two.}
    \label{fig:modulesVariables}
\end{figure}
\vspace{6pt}
    As an example, let us take the code for module 1 and module 2 from Listing~\ref{module1and2} and the following requirement:

\vspace{.2cm}
\begin{enumerate}[label=(\subscript{R}{{\arabic*}})]\setcounter{enumi}{2}
    \item Always, at the end of the execution of module 2, \texttt{v\_2}=\texttt{TRUE}.
\end{enumerate}
\vspace{.1cm}

\begin{lstlisting}[caption={Module 1.}, style=assertion, label=module1and2]
FUNCTION_BLOCK module_1
    VAR_OUTPUT
        v_1 : BOOL;
    END_VAR
BEGIN
    v_1 := FALSE;
END_FUNCTION_BLOCK

FUNCTION_BLOCK module_2
    VAR_INPUT
        v_1 : BOOL;
    END_VAR
    VAR_OUTPUT
        v_2 : BOOL;
    END_VAR
BEGIN
    v_2 := NOT v_1;
END_FUNCTION_BLOCK

\end{lstlisting}

    If we only verify module 2 for every possible value of \texttt{v\_1}, we would get the counterexample \{\texttt{v\_1}=\texttt{TRUE}, \texttt{v\_2}=\texttt{FALSE}\}. However, \texttt{v\_1} only takes the value \texttt{FALSE} at the end of module 1, which is the input of module 2. Therefore, this counterexample is not real.

    An option to specify this requirement is shown below. We have the assumption from $R_4$ and the conditional requirement $R'_3$ based on this assumption. Now, no counterexamples would be found.

\vspace{.2cm}
\begin{enumerate}[label=(\subscript{R}{{\arabic*}})]\setcounter{enumi}{3}
    \item Always, at the end of the execution of module 1, \texttt{v\_1}=\texttt{FALSE}.
\end{enumerate}
\begin{enumerate}[label=(\subscript{R'}{{\arabic*}})]\setcounter{enumi}{2}
    \item Given that \texttt{v\_1}=\texttt{FALSE} at the beginning of module 2, always, at the end of the execution of module 2, \texttt{v\_2}=\texttt{TRUE}.
\end{enumerate}
\vspace{.2cm}

    Nevertheless, it is important to note that modules are recommended to be robust to any possible input values. It can happen that, due to, e.g., hardware failures, variables take other values that were not supposed to take.
    
    \item \textbf{Timers}. When time is involved in the system, formalizing requirements becomes harder and more error-prone. In this case, every step needs to be formalized, such as when timers are activated, what happens before reaching the total time, what happens afterward, how it is reset, etc.

    Figure~\ref{fig:modesTimer} shows a state machine in which it is possible to transition from \texttt{mode 1} to \texttt{mode 2} if $\varphi$ is true after a certain amount of time. However, it is not specified if and how the timer is reset and what happens if $\varphi$ is not true.

    On the other hand, in Figure~\ref{fig:modesTimerAutomaton}, we have created a timed automaton specifying how the timer (clock) works. It can also be reset if $\psi$ is true.
    
\end{itemize}

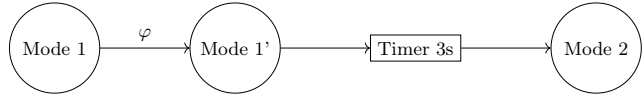
\begin{figure}[H]
    \centering
    \scalebox{.8}{\begin{tikzpicture}[->, node distance=3cm]

  \node [state, align=center, minimum width=1.5cm] (s0) {Mode 1};
  \node [state, align=center, minimum width=1.5cm] (s0wait) [right of=s0] {Mode 1'};
  \node [draw, shape=rectangle, align=center, minimum width=1.5cm] (timer) [right of=s0wait] {Timer 3s};
  \node [state, align=center, minimum width=1.5cm] (s1) [right of=timer] {Mode 2};
  
  \path[every node/.style={font=\sffamily\small}]
    (s0.east) edge node [above, align=center] {$\varphi$} (s0wait.west) 
    (s0wait.east) edge node [above, align=center] {} (timer.west)
    (timer.east) edge node [above, align=center] {} (s1.west)
  
    ;
        
\end{tikzpicture}}
    \caption{Example of diagram with ambiguous timer.}
    \label{fig:modesTimer}
\end{figure}
\vspace{12pt}
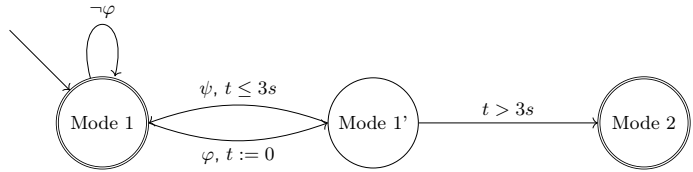
\begin{figure}[H]
    \centering
    \scalebox{.8}{\begin{tikzpicture}[->, node distance=4.5cm]

  \node [state, accepting, align=center, minimum width=1.5cm] (s0) {Mode 1};
  \node [] (init) [above left=1cm and 1cm of s0] {};
  \node [state, align=center, minimum width=1.5cm] (s0wait) [right of=s0] {Mode 1'};
  \node [state, accepting, align=center, minimum width=1.5cm] (s1) [right of=s0wait] {Mode 2};
  
  \path[every node/.style={font=\sffamily\small}]
    (init) edge node {} (s0)
    (s0.east) edge  [bend right=20] node [below, align=center] {$\varphi$, $t:=0$} (s0wait.west) 
    (s0wait.east) edge node [above, align=center] {$t>3s$} (s1.west)
    (s0wait.west) edge [bend right=20] node [above, align=center] {$\psi$, $t\leq 3s$} (s0.east)

    (s0) edge [loop above] node [above, align=center] {$\neg\varphi$} (s0)
    ;
        
\end{tikzpicture}}
    \caption{Example of timed automaton representing the use of a timer.}
    \label{fig:modesTimerAutomaton}
\end{figure}

\fi

\end{document}